\documentclass[letterpaper, 10 pt, conference]{ieeeconf}  % Comment this line out if you need a4paper

\IEEEoverridecommandlockouts                              % This command is only needed if 
                                                          \usepackage[utf8]{inputenc}
\usepackage{blindtext}
\usepackage{graphicx}
\usepackage[utf8]{inputenc}
\usepackage{graphics} % for pdf, bitmapped graphics files
\usepackage{epsfig} % for postscript graphics files
\usepackage{mathptmx} % assumes new font selection scheme installed
\usepackage{amsmath} % assumes amsmath package installed
\usepackage{amssymb}  % assumes amsmath package installed
\usepackage{cite}
\usepackage{multicol}
\usepackage{mathtools, cuted}
\usepackage[cmintegrals]{newtxmath}
\usepackage{epstopdf}
\usepackage{graphics} % for pdf, bitmapped graphics files
\usepackage{epsfig} 

\usepackage{titlesec}
\usepackage{epsfig} % for postscript graphics files
\usepackage{mathptmx} % assumes new font selection scheme installed
\usepackage{amsmath} % assumes amsmath package installed
\usepackage{amssymb}  % assumes amsmath package installed
\usepackage{cite}
\usepackage{multicol}
\usepackage{mathtools, cuted}
\usepackage[cmintegrals]{newtxmath}
\usepackage{makeidx}         % allows index generation
\usepackage{algorithmic}

\usepackage{multicol}        % used for the two-column index
\usepackage[bottom]{footmisc}% places footnotes at page bottom
\usepackage{caption}
\usepackage{nomencl}
\usepackage[usenames, dvipsnames]{color}

 \usepackage{amsthm}
 \theoremstyle{definition}

\theoremstyle{theorem}
\newtheorem{theorem}{Theorem}

\theoremstyle{assumption}
\newtheorem{assumption}{Assumption}

\theoremstyle{lemma}

\theoremstyle{remark}
\newtheorem{remark}{Remark}

 \theoremstyle{proposition}

\usepackage{algorithmic}    % Include this line if your 
\usepackage[ruled,vlined]{algorithm2e}

\usepackage{array}
\usepackage{float}
\usepackage{subcaption}
\usepackage{graphicx}
\title{\LARGE \bf
Boundary Control of Traffic Congestion Modeled as a Non-stationary Stochastic Process
}

\author{Xun Liu and Hossein Rastgoftar% <-this % stops a space
\thanks{Xun Liu is with the Department of Mechanical Engineering at Villanova University, Villanova, PA 19085, USA {\tt\small xliu8@villanova.edu}}
\thanks{Hossein Rastgoftar is with the Department of Aerospace and Mechanical Engineering at the University of Arizona, Tucson, AZ 85721, USA {\tt\small hrastgoftar@arizona.edu}}
}

\begin{document}

\maketitle
\thispagestyle{empty}
\pagestyle{empty}

%%%%%%%%%%%%%%%%%%%%%%%%%%%%%%%%%%%%%%%%%%%%%%%%%%%%%%%%%%%%%%%%%%%%%%%%%%%%%%%%
\begin{abstract}
In this paper, we introduce a new conservation-based approach to model traffic dynamics, and apply the model predictive control (MPC) approach to manage the boundary traffic inflow and outflow, so that the traffic congestion is reduced. We establish an interface between the Simulation of Urban Mobility (SUMO) software and MATLAB to define a network of interconnected roads (NOIR) as a directed graph, and present traffic congestion management as a network control problem. By formally specifying the traffic feasibility conditions, and using the linear temporal logic, we present the proposed MPC-based boundary control problem as a quadratic programming with linear equality and inequality constraints. The success of the proposed traffic boundary control is demonstrated by simulation of traffic congestion control in Center City Philadelphia.

\end{abstract}

%%%%%%%%%%%%%%%%%%%%%%%%%%%%%%%%%%%%%%%%%%%%%%%%%%%%%%%%%%%%%%%%%%%%%%%%%%%%%%%%
\section{Introduction}
During the urbanization process, numerous negative impacts have been created by the traffic congestion, such as environmental pollution  \cite{chin2011impact}, economic recession  \cite{liang2013road,annan2015traffic}, human physical and mental health harms  \cite{muneera2018economic}, and ecological destruction \cite{margaret2004impact}. Considering the acceleration of the urbanization process, it is urgent to solve the traffic congestion problem. Some researchers have presented several temporary solutions to reduce traffic congestion, such as building more roads or restricting  vehicles; however, these solutions are not in consideration in this paper. We concentrate on finding the optimal solution by constructing and controlling the traffic model. 

% With the rapid development of urbanization and the popularization of private vehicles, the problem of urban traffic congestion has become more and more prominent. Congestion can lead to numerous negative impacts. It will not only leads to the decline of economic and social well-being \cite{muneera2018economic, margaret2004impact}, but also cause the continuous deterioration of the urban living environment which can hinder the development of the society  \cite{liang2013road, annan2015traffic}. Furthermore, traffic congestion can destroy the urban environment and ecology. Due to the low-speed driving condition, the emission of greenhouse gas, noxious gas and noise will increase and that will badly affect human health  \cite{chin2011impact}. 
Over the past years, researchers have proposed a large number of approaches to alleviate traffic congestion. We can divide the approaches into two categories: model-based approaches and model-free approaches. While the model-based approaches rebuild traffic dynamics properties in a traffic model and obtain the optimal solution by applying the appropriate control method to the traffic model, the model-free approaches accomplish the traffic management by replacing the traffic model through an equivalent data model or controlling of the traffic signal plan.   

An essential task in the model-based approach is to construct a virtual traffic model to represent the real traffic dynamics' properties. Then, based on this traffic model, appropriate approaches can be applied to control the traffic dynamics or predict the traffic states. The macroscopic fundamental diagram (MFD) of the traffic flow model \cite{geroliminis2008existence}, which describes the relationship between traffic density and traffic flow, is a widely used approach to describe the traffic dynamics. Ref. \cite{xu2013traffic} verifies the MFD model based on the real traffic data and evaluates the traffic state by applying this model. Moreover, Ref. \cite{sirmatel2017integration} integrates the perimeter control with the MFD model to improve the traffic network capacity and mobility. To improve the accuracy of the traffic dynamics model, Ref.  \cite{shao2018distributed,munoz2003traffic,yin2017offblock,feldman2002cell,yang2017fundamental,yang2019network} integrate the cell transmission model (CTM) approach with the MFD model. Furthermore, the CTM approach, which is widely used to partition the traffic network into road elements, can also incorporate conservation laws to construct the traffic dynamics model  \cite{liu2021conservation,rastgoftar2019integrative,zhang2021optimization,ba2016ondistributed}. Moreover, Ref. \cite{rastgoftar2021physicsbased} incorporates the CTM model with the finite-state Markov decision process (MDP) to obtain the optimal movement phases at the traffic junctions. Once the traffic model is generated, the control and optimization methods can be implemented to find the optimal solution. Ref. \cite{sirmatel2017integration,rastgoftar2019integrative,rastgoftar2020resilient,rastgoftar2021physicsbased} adopt the model predictive control (MPC) approach to obtain the optimal solution for boundary traffic control and Ref. \cite{lin2011fast,zhang2021optimization} apply mixed-integer linear programming (MILP) to solve the optimization problem.

Traffic congestion control can also be accomplished through a model-free approach. For example, Ref. \cite{li2019model-free} uses a data model to represent the traffic dynamics and incorporates adaptive predictive control to adjust the boundary input. Another model-free possibility for reducing congestion is through traffic signal optimization. In recent years, with the rapid development of AI technology and computing capacity, the reinforcement learning (RL) has been increasingly adopted to optimize the traffic signal plan. Ref.  \cite{abdulhai2003reinforcementlearning} proposes a Q-learning algorithm for traffic signal control. Moreover, Ref. \cite{prashanth2011reinforcementlearning} integrates the RL with the Markov decision process to reduce computational complexity. Furthermore, Ref. \cite{lin2018efficient} introduces an efficient RL method for traffic signal control in complicated multiple intersections and Ref. \cite{greguric2020application} summarizes the RL algorithms that have been applied in adaptive traffic signal control (ATSC) in recent years. In addition to the RL approach, a fuzzy controller is also employed by researchers to control the traffic signal \cite{oluyemi2019development,bhatia2021smart}.

\begin{figure}
    \centering
    \includegraphics[width=0.48\textwidth]{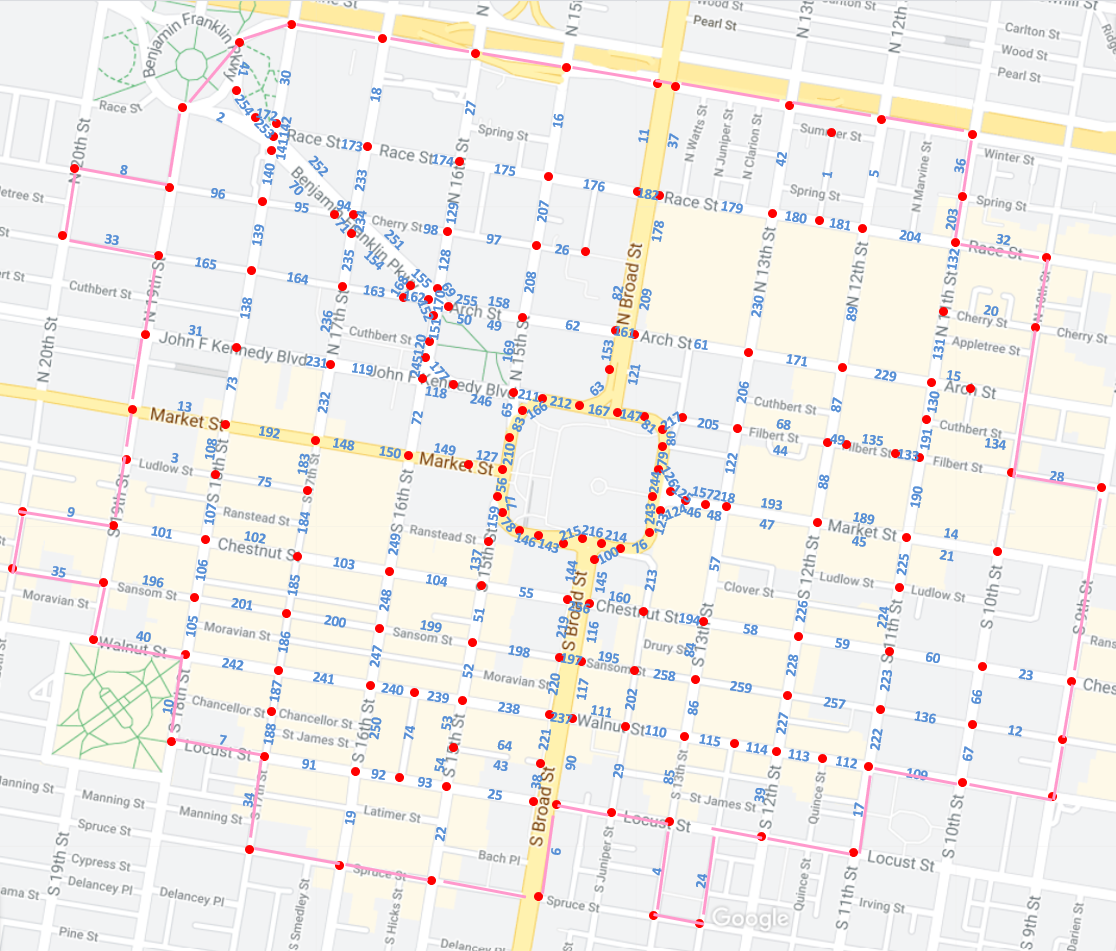}
    \caption{Example NOIR: Center City, Philadelphia}
    \label{fig:Philadelphia Center}
\end{figure}

This paper presents further research based on our previous work \cite{liu2021conservation}. In this paper, we continue to use mass conservative law to define the traffic coordination model and describe traffic dynamics. Also, we model the dynamics of the traffic coordinate as a non-stationary stochastic process. Comparing with our previous research, in this paper we modify the definition of the stochastic parameter to improve and perfect the traffic model to make it more realistic. In addition, in the process of determining the optimal boundary control solution, we consider and simulate the impact of the current number of road vehicles in the results. Similar to our previous work, the network of interconnected roads (NOIR) is generated by converting the real street map using the software Simulation of Urban Mobility (SUMO) and MATLAB. The road elements in the NOIR are categorized into three groups: "inlet road elements," "outlet road elements," and "interior road elements." We set the boundary traffic flow as the control variable, and accomplish the objective of traffic congestion alleviation by controlling the boundary inflow from the inlet road elements and the boundary outflow from the outlet road elements using the model predictive control (MPC) approach. In the case study, we integrate the proposed traffic model with the MPC approach and illustrate the results of traffic congestion management on the Center City district of Philadelphia (See. Fig. \ref{fig:Philadelphia Center}).

This paper is organized as follows: Section \ref{section 2} gives an introduction about the notions existing in the NOIR and the linear temporal logic symbols used to describe traffic state conditions. The basic principles and definitions for model construction and the traffic feasibility conditions for solving the cost function in boundary control are explained in Section \ref{section 3}. The explanation about traffic network dynamics and the traffic control approach are presented in Section \ref{section 4} and Section \ref{section 5}, respectively. The simulation results of a case study using the proposed traffic model and control approach in Center City Philadelphia  are illustrated in Section \ref{section 6}, followed by the conclusion in Section \ref{section 7}.

\section{Preliminaries}\label{section 2}
\subsection{Graph Theory Notions}
The network of inter-connected roads (NOIR) presented in this paper is generated by using the software Simulation of Urban Mobility (SUMO) based on a real street map of Philadelphia. A NOIR can be denoted as a graph $\mathcal{G}\left(\mathcal{V},\mathcal{E}\right)$, where set $\mathcal{V}=\left\{1,\cdots,N\right\}$ defines $N$ road elements and set $\mathcal{E}\subset \mathcal{V}\times \mathcal{V}$ determines  the interconnections between the road elements. As shown in Fig.  \ref{fig:Philadelphia Center}, every element $i\in\mathcal{V}$ is a unidirectional street and located between two consecutive junctions. Element $\left(i,j\right)\in\mathcal{E}$ represents the connection directed from road element $i\in\mathcal{V}$ to road element $j\in\mathcal{V}$. 
Set $\mathcal{V}$ can be divided into three subsets: the inlet road elements set $\mathcal{V}_{in}=\{1,\cdots,N_{in}\}$, the outlet road elements set $\mathcal{V}_{out}=\{N_{in}+1,\cdots,N_{out}\}$, and the interior road elements set $\mathcal{V}_{I}=\{N_{out}+1,\cdots,N\}$.  For every road element $i\in \mathcal{V}$, we define in-neighbor set $\mathcal{I}_{i}$ and out-neighbor set $\mathcal{O}_{i}$ as follows:
\begin{subequations}
\begin{equation}
    \mathcal{I}_i=\{j|\left(j,i\right)\in \mathcal{E}\}{\color{black},}
\end{equation}
\begin{equation}
    \mathcal{O}_i=\{j|\left(i,j\right)\in \mathcal{E}\}{\color{black},}
\end{equation}
\end{subequations}
where the in-neighbor and out-neighbor road elements refer to the upstream adjacent road elements and downstream adjacent road elements. Note that the in-neighbor set of every inlet road element and the out-neighbor set of every outlet road element are empty, i.e., $\mathcal{I}_i=\emptyset$, if $i\in\mathcal{V}_{in}$ and $\mathcal{O}_i=\emptyset$, if $i\in\mathcal{V}_{out}$.

\subsection{Linear Temporal Logic}
We  use the linear temporal logic (LTL) to describe the properties and feasible conditions of the traffic dynamics model. A LTL formula normally consists of three components, the propositional variables, the logical operators, and the temporal modal operators\cite{zohar1991temporal}. Propositional variable is the most fundamental element in the propositional logic whose value is either true or false. The logical operators, such as \textit{negation} $\left(\lnot\right)$, \textit{disjunction} $\left(\wedge\right)$ and \textit{conjunction} $\left(\vee\right)$, can act on a single propositional variable or between multiple propositional variables to express a sophisticated logical formula accurately. The temporal modal operators including  \textit{eventually} $\left(\Diamond\right)$, \textit{always} $\left(\Box\right)$, \textit{next} $\left(\bigcirc\right)$, and \textit{until} $\left(\mathcal{U}\right)$, define the temporal variables of  LTL formulas \cite{wongpiromsarn2009receding}. 

Inspired by Metric Temporal Logic (MTL) \cite{koymans1990specifying,rastgoftar2019integrative}, we extend the classic LTL  by integrating the temporal modal operators with a distance function to restrict the logical formula into a finite domain. For example, assuming $\mathcal{T}=\{0,\cdots,N_\tau| 0<N_{\tau}<\infty\}$ is a finite time domain, the formula $\Box_{\mathcal{T}}\left(\cdot\right)$ at sampling time $k$ means that the statement expressed in the parenthesis is always true at every time $k+t$, where $t\in\mathcal{T}$.
% Those subsets can be denoted as $\mathcal{V}_{in}=\{1,\cdots,N_{in}\}$, $\mathcal{V}_{out}=\{N_{in}+1,\cdots,N_{out}\}$ and $\mathcal{V}_{I}=\{N_{out}+1,\cdots,N\}$ respectively. The relationship between the road elements sets is: $\mathcal{V}=\mathcal{V}_{in}\bigcup \mathcal{V}_{out}\bigcup \mathcal{V}_{I}$. The interaction between road elements are denoted with 

% A network of inter-connected roads {\color{black}(NOIR) is represented by graph $\mathcal{G}\left(\mathcal{V},\mathcal{E}\right)$ with node set $\mathcal{V}$ and edge set $\mathcal{E}\subset \mathcal{V}\times \mathcal{V}$}. {\color{black}Node} $i\in\mathcal{V}$ {\color{black}represents an NOIR road, and edge $\left(i,j\right)\in \mathcal{E}$ represents a directed connection from road $i\in \mathcal{V}$ to road $j\in \mathcal{V}$.} 
% {\color{black}Set} $\mathcal{V}$ can be partitioned as $\mathcal{V}=\mathcal{V}_{in}\bigcup \mathcal{V}_{out}\bigcup \mathcal{V}_{I}$, where denote index numbers of inlet, outlet, and interior road elements. 
% {\color{black}Sets
% \begin{subequations}
% \begin{equation}
% \mathcal{I}_i=\left\{j:\left(j,i\right)\in \mathcal{E}\right\},
% \end{equation}
% \begin{equation}
% \mathcal{O}_i=\left\{j:\left(i,j\right)\in \mathcal{E}\right\}
% \end{equation}
% \end{subequations}
% define in-neighbors and out-neighbors of road $i\in \mathcal{V}$. 

% }

\section{Problem Statement}\label{section 3}
% Figure   \ref{fig:Figure 1} illustrate the traffic flow in the network.

% \begin{figure} [H]
%     \centering
%     \includegraphics[width=0.8\textwidth]{UpdatedTrafficFlow.png}
%     \caption{Traffic flow sketch in the network}
%     \label{fig:Figure 1}
% \end{figure}
In this paper, we consider the traffic dynamics as a physical model which satisfies the mass conservation law and describe the dynamics in every road element $i\in\mathcal{V}$ as follows:
\begin{equation}\label{traffic dynamics genreal}
    \rho_{i}[k+1]=\rho_{i}[k]+y_{i}[k]-z_{i}[k]+s_{i}[k]
\end{equation}
where $k=1,2,\cdots$ denotes the discrete sampling time, $\rho_{i}[k]$ is the number of existing cars at road element $i$, and called \textit{traffic density}. Also. $y_i[k]$, $z_i[k]$ and $s_i[k]$ are the network traffic inflow, network traffic outflow and the external traffic flow, respectively. 

The external traffic flow $s_i[k]$, which is defined in Eq. \eqref{external flow}, specifies the traffic exchange between the NOIR and the external environment. It is prescribed that, at every sampling interval $[t_{k},t_{k+1})$ the traffic could only drive into the NOIR through the inlet road elements, defined by  $\mathcal{V}_{in}$, and depart from the NOIR through the outlet road elements, defined by  $\mathcal{V}_{out}$.  Therefore, we define $s_i[k]$ by
\begin{equation}\label{external flow}
    s_{i}[k]=
    \begin{cases}
    u_{i}[k] \geq 0  & {i\in \mathcal{V}_{in}}\\
    -v_{i}[k] \leq 0 & {i\in \mathcal{V}_{out}}\\
    0                & {i\in \mathcal{V}_I}
    \end{cases}
\end{equation}
at every discrete time $k$, where $u_i[k]\geq 0$ is the number of cars entering the NOIR through $\mathcal{V}_{in}$ and $v_i[k]\geq 0$ is the number of cars leaving the NOIR through $\mathcal{V}_{out}$.

\begin{figure}
\centering
     \begin{subfigure}{.15\textwidth}
         \centering
         \includegraphics[width=\textwidth]{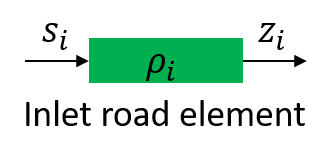}
         \caption{Inlet road element $i\in\mathcal{V}_{in}$}
         \label{fig:flow inlet road}
     \end{subfigure}
     \hfill
     \begin{subfigure}{.15\textwidth}
         \centering
         \includegraphics[width=\textwidth]{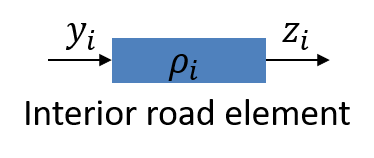}
         \caption{Interior road element $i\in\mathcal{V}_{I}$}
         \label{fig:flow interior road}
     \end{subfigure}
     \hfill
     \begin{subfigure}{.15\textwidth}
         \centering
         \includegraphics[width=\textwidth]{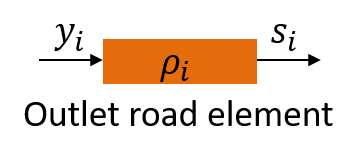}
         \caption{Outlet road element $i\in\mathcal{V}_{out}$}
         \label{fig:flow outlet road}
     \end{subfigure}
    \caption{Traffic flows in inlet road element, interior road element and outlet road element}
    \label{Traffic flow of road in NOIR}
\end{figure}

The network traffic inflow $y_i[k]$ and traffic outflow $z_i[k]$ given by Eq. \eqref{network inflow and outflow} determine the traffic flow exchange between road elements within the NOIR at every discrete sampling time $k$, and defined by
\begin{subequations}\label{network inflow and outflow}
\newlength{\widest}
\settowidth{\widest}{$v_{i}[k]=q_{i,j}[k]z_{j}[k]$}
\begin{align}
    y_{i}[k]=&
    \begin{cases}\label{network inflow}
    0&\quad i\in\mathcal{V}_{in}\\
    v_{i}[k]&\quad i\in\mathcal{V}_{out}\\
    \sum_{j\in\mathcal{I}_{i}}q_{i,j}[k]z_{j}[k]&\quad i\in\mathcal{V}_{I}
    \end{cases}{\color{black},}\\
    z_{i}[k]=&
    \begin{cases}\label{network outflow}
    \makebox[\widest][l]{$u_{i}[k]$}&\quad i\in\mathcal{V}_{in}\\
    0&\quad i\in\mathcal{V}_{out}\\
    p_{i}[k]\left(\rho_{i}[k]+y_i[k]\right)&\quad i\in\mathcal{V}_{I}
    \end{cases}
    ,
\end{align}
\end{subequations}
where
\begin{equation}
\label{outflowprobability}
    p_i[k]=\begin{cases}
    1&\mathrm{If~}i\in \mathcal{V}_{in}\bigcup\mathcal{V}_{out}\\
    0&\mathrm{If~}i\in \mathcal{V}_{I}~\mathrm{and}~y_i+\rho_i=0\\
    {z_i[k]\over \rho_i[k]+y_i[k]}&\mathrm{If~}i\in \mathcal{V}_{I}~\mathrm{and}~y_i+\rho_i\neq0\\
    \end{cases}
\end{equation}
is the fraction of inflow and outflow of road $i\in\mathcal{V}$ that leaves $i\in \mathcal{V}$  during sampling interval $[t_{k},t_{k+1})$. Also, $q_{i,j}[k]\in [0,1]$ in Eq. \eqref{network inflow} is the fraction of cars driving from road element $j\in\mathcal{V}\setminus\mathcal{V}_{out}$ to each of its downstream adjacent road elements $i\in\mathcal{V}$ at every sampling time $k$. Therefore, $q_{i,j}[k]$ must satisfy the following equality constraint:
\begin{equation}
    \sum_{i\in O_{j}} q_{i,j}= 1.
\end{equation}
Note that the flow probability $p_i[k]$ and fraction probability $q_{i,j}[k]$ reflect the uncertainty caused by human driver intentions. Therefore, our proposed model consistently incorporates the human intent into modeling of traffic coordination. In this paper, we assume that the flow probability $p_i[k]$ and fraction probability $q_{i,j}[k]$ are known but they are randomly generated at every road $i\in \mathcal{V}$ and every discrete time $k=0,1,2,\cdots$. Fig.\ref{Traffic flow of road in NOIR} illustrates the traffic flow at inlet road element, interior road element and outlet road element.

\begin{assumption}\label{assumption 1}
In the process of building the traffic dynamics model, we assume that the time increment $\Delta T=t_{k+1}-t_{k}$ is constant and sufficiently small such that the flow probability $p_{i}[k]$  satisfies the following condition:
\begin{equation}
    p_i[k]=
    \begin{cases}
    \dfrac{z_i[k]}{\rho_i[k]+y_i[k]}\in \left[0,1\right)&\mathrm{If} ~\rho_i[k]+y_i[k]\neq 0\\
    0&\mathrm{If}~\rho_i[k]+y_i[k]= 0\\
    \end{cases}
    .
\end{equation}
% at every discrete sampling time $k$,if $\rho_i[k]+y_i[k]\neq 0$. Moreover, as defined in Eq.$ \eqref{outflowprobability}$, the $p_{i}[k]=0$ over the time interval $\left[t_{k},t_{k+1}\right)$, if $\rho_i[k]+y_i[k]= 0$.
\end{assumption}

% {\color{black}\begin{assumption}\label{assumption1}
% Discrete time $k$ represents time interval $\left[t_{k},t_{k+1}\right)$, and time increment $\Delta T=t_{k+1}-t_k$ is constant for $k=0,1,2,\cdots$. We choose a sufficiently-small time increment $\Delta T$ such that
% \begin{equation}
%     p_i[k]=\dfrac{z_i[k]}{\rho_i[k]+y_i[k]}\in \left[0,1\right)
% \end{equation}
% at every road $i\in \mathcal{V}_I$ and every discrete time $k$ if $\rho_i[k]+y_i[k]\neq 0$. Note that $p_i[k]=0$, if $\rho_i[k]+y_i[k]=0$ over the time interval $t\in \left[t_k,t_{k+1}\right)$ (See Eq.  \eqref{outflowprobability}).
% \end{assumption}}

In this paper, we offer an MPC-based boundary control  to manage the traffic congestion in the NOIR. Therefore, boundary input $s_i$ is determined at every road $i\in \mathcal{V}_{in}\bigcup \mathcal{V}_{out}$ by solving  a quadratic programming problem with the cost function and constraints that are described below:

\textbf{Traffic Coordination Cost:} We define
\begin{equation}\label{cost function}
    \begin{split}
    \mathrm{C}=&{1\over 2}\sum_{j=k+1}^{k+N_\tau}\left(\sum_{i\in \mathcal{V}_{in}\bigcup\mathcal{V}_{out}}s_i^2[j-1]+\beta\sum_{l\in \mathcal{V}_{I}}\rho_l^2[j]\right)\\
    % =&\min_{i_{1},\cdots,i_{N_{in}};j_{N_{in}+1},\cdots,j_{N_{out}}}{1\over 2}\left(\sum_{i\in \mathcal{V}_{in}}u_i^2[k]+\sum_{j\in \mathcal{V}_{out}}v_j^2\left[k\right]\right)
\end{split}
\end{equation}
as the traffic coordination cost, where
$s_{i}=u_{i}$ for every inlet road element $i\in\mathcal{V}_{in}$, $s_{i}=v_{i}$ for every outlet road element $i\in\mathcal{V}_{out}$, and $\beta>0$ is a constant scaling parameter, which reflects the influence weight of the number of existing cars on the interior road element $\rho_l$ on the criterion function $\mathrm{C}$.

% , and $\rho_{i}$ represent the traffic density at every interior road element $i\in\mathcal{V}_{I}$. Note that, at every sampling time $k$ the traffic density $\rho_{i}[k]$ is known and the density variable $\rho_{i}[j]$ is a function with respect to $\rho_{i}[k]$ and $s_{i}[j-1]$, i.e., $\rho_{i}[j]=f(\rho_{i}[k],s_{i}[j-1])$.
% Note that cost function $\mathrm{C}$ $s_i[j-1]$ and $\rho_l[j-1]$, where $\rho_l[k]$ is known and denotes the traffic density of the interior road element $l\in\mathcal{V}_{I}$ at every discrete time $k$.
% the cost function $\mathrm{C}$ could be formulated as: 
% $\mathrm{C}=\mathrm{C}\left(\rho_{i}[k],s_{i}[k],\cdots,s_{i}[k+N_{\tau}-1]\right)$. 
% Meanwhile, in order to obtain the optimal solutions, the above cost function $\mathrm{C}$ is subject to the following traffic feasibility conditions:

\textbf{State Feasibility Condition:} Traffic density  is set to be a non-negative physical parameter. Moreover, we assume every road element in the NOIR holds a maximal capacity $\rho_{max}$ and the traffic density of the road element can not exceed this maximal value within the next $N_\tau$ time steps at every sampling time $k$. We define a LTL formula $\pi_{1}$ to express these state constrains at every interior road element $i\in\mathcal{V}_{I}$ by 
\begin{equation}\label{State Feasibility Condition}
    \pi_{1}\coloneqq\bigwedge_{i\in\mathcal{V_{I}}}\Box_{\mathcal{T}}\left(\rho_{i}>0\wedge\rho_{i}<\rho_{max}\right).
\end{equation}
% \begin{equation}
%     \pi_{1}\coloneqq\bigwedge_{k\in\mathcal{K}}\bigwedge_{i\in\mathcal{V_{I}}}\left(\rho_{k,i}>0\wedge\rho_{k,i}<\rho_{max}\right)
% \end{equation}

\textbf{Input Feasibility Condition:} At every sampling time $k$, back-flow must be prohibited at every boundary road element $i\in\mathcal{V}_{in}\bigcup\mathcal{V}_{out}$ within the next $N_\tau$ sampling times. We express this feasibility condition using the LTL formula $\pi_{2}$:
\begin{equation}\label{Input Feasibility Condition}
    \pi_{2}\coloneqq\bigwedge_{i\in\mathcal{V}_{in}}\bigwedge_{j\in\mathcal{V}_{out}}\Box_{\mathcal{T}}\left(u_{i}\geq0\wedge v_{j}\geq0\right)
\end{equation}

\textbf{Input Optional Condition:} We assume that the demand for entering and leaving the NOIR is high and only $d_{0}$ amount of cars are permitted to cross the boundary of the NOIR within the next $N_\tau$ sampling times at every discrete time $k$. Therefore, the following LTL formula must be satisfied:
\begin{equation}\label{Input Optional Condition}
    \pi_{3}\coloneqq\Box_{\mathcal{T}}\left(\sum_{i\in\mathcal{V}_{in}}u_{i}+\sum_{j\in\mathcal{V}_{out}}v_{j}=d_{0}\right)
\end{equation}

\section{Traffic Network Dynamics}\label{section 4}
We substitute Eqs. \eqref{external flow} and  \eqref{network inflow and outflow} into Eq. \eqref{traffic dynamics genreal} and simplify the traffic dynamics for every road element $i\in\mathcal{V}$ as follows:

\begin{subequations}\label{traffic dynamics}
\begin{equation}\label{traffic dynamics boundary node}
    \forall i\in\mathcal{V}_{in}\bigcup\mathcal{V}_{out},\quad\rho_i[k+1]=\rho_i[k]
\end{equation}
\begin{equation}\label{traffic dynamics interior node}
\resizebox{0.99\hsize}{!}{%
$
    \forall i\in\mathcal{V}_{I}, \rho_i[k+1]=\left(1-p_i[k]\right)\left(\rho_i[k]+\sum_{j\in \mathcal{I}_i}q_{i,j}[k]z_j[k]\right).
$
}
\end{equation}
\end{subequations}

Eq. \eqref{traffic dynamics} implies that the traffic density remains constant at every road $i\in \mathcal{V}_{in}$ but it is updated with time at every interior road elements $i\in\mathcal{V}_{I}$. As a result, the network traffic dynamics are only defined over the interior road elements. To obtain the traffic dynamics, we define the state vector $\mathbf{x}=
    \begin{bmatrix}
    \rho_{N_{out}+1}&\cdots&\rho_N
    \end{bmatrix}^\mathsf{T}\in \mathbb{R}^{\left(N-N_{out}\right)\times 1}$, the inflow vector $\mathbf{y}\in \mathbb{R}^{\left(N-N_{out}\right)\times 1}$, and the outflow vector $\mathbf{z}\in \mathbb{R}^{\left(N-N_{out}\right)\times 1}$. 
Moreover, we define the outflow probability matrix $\mathbf{P}\in\mathbb{R}^{\left(N-N_{out}\right)\times \left(N-N_{out}\right)}$ and the tendency probability matrix $\mathbf{Q}\in\mathbb{R}^{\left(N-N_{out}\right)\times \left(N-N_{out}\right)}$ as follows:
\begin{subequations}
\begin{equation}
    \mathbf{P}\left[k\right]=\mathrm{diag}\left(p_{N_{out}+1}\left[k\right],\cdots,p_N\left[k\right]\right),
\end{equation}
\begin{equation}
    \mathbf{Q}\left[k\right]=\left[Q_{ij}\left[k\right]\right]=\left[q_{i+N_{out},j+N_{out}}\left[k\right]\right],
\end{equation}
\end{subequations}
% \begin{equation}
% \begin{split}
%     \\
%     \mathbf{Q}\left[k\right]&=\left[Q_{ij}\left[k\right]\right]=\left[q_{i+N_{out},j+N_{out}}\left[k\right]\right]
% \end{split}
% \end{equation}
where $Q_{ij}=q_{i+N_{out},j+N_{out}}$ determines the fraction of departing vehicles driving from road element $\left(j+N_{out}\right)\in \mathcal{V}_I$ towards road element $\left(i+N_{out}\right)\in \mathcal{V}_I$ at discrete time $k$.

 By considering definitions of traffic inflow and outflow given in Eq.  \eqref{network inflow and outflow}, the network inflow vector $\mathbf{y}$ and network outflow vector $\mathbf{z}$ are related to state vector $\mathbf{x}$  by
\begin{subequations}
\begin{equation}\label{law_y}
    \mathbf{y}[k]= \left(\mathbf{I}-\mathbf{Q}[k]\mathbf{P}[k]\right)^{-1}\mathbf{Q}[k]\mathbf{P}[k]\mathbf{x}[k],
\end{equation}
\begin{equation}
\label{law_z}
    \mathbf{z}[k]=\left(\mathbf{P}[k]\left(\mathbf{I}-\mathbf{Q}[k]\mathbf{P}[k]\right)^{-1}\mathbf{Q}[k]\mathbf{P}[k]+\mathbf{P}[k]\right)\mathbf{x}[k],
\end{equation}
\end{subequations}
 at every sampling time $k$, and 
% Since Eq. \eqref{traffic dynamics interior node} is applied to model the traffic coordination for every interior road element $i\in\mathcal{V}_{I}$, 
the traffic dynamics can be expressed in the state space form  by 
\begin{equation}\label{traffic dynamic state vector}
    \mathbf{x}[k+1]=\mathbf{A}[k]\mathbf{x}[k]+\mathbf{B}[k]\mathbf{s}[k].
\end{equation}
In Eq. \eqref{traffic dynamic state vector}, $\mathbf{s}[k]=[s_{i}[k]]\in\mathbb{R}^{N_{out}\times1}$, $\mathbf{B}[k]=b_{i,j}[k]\in\mathbb{R}^{\left(N-N_{out}\right)\times N_{out}}$, and $\mathbf{A}[k]\in\mathbb{R}^{\left(N-N_{out}\right)\times \left(N-N_{out}\right)}$ refer to the input vector, the input matrix, and the system matrix at every sampling time $k$, where
\begin{subequations}
\begin{equation}
    s_{i}[k]=
    \begin{cases}
    u_{i}[k],&\mathrm{If~} i\in\mathcal{V}_{in}=\{1,\cdots,N_{in}\}\\
    v_{i}[k],&\mathrm{If~} i\in\mathcal{V}_{out}=\{N_{in}+1,\cdots,N_{out}\}
    \end{cases},\\
\end{equation}
\begin{equation}
    b_{ij}[k]=
    \begin{cases}
    1\quad j\in\mathcal{I}_{i+N_{out}}\\
    -1\quad j\in\mathcal{O}_{i+N_{out}}
    \end{cases},\\
\end{equation}
\begin{equation}
\begin{split}
    \mathbf{A}[k]=&\left(\mathbf{I}-\mathbf{P}[k]\right)\left(\mathbf{I}+\left(\mathbf{I}-\mathbf{Q}[k]\mathbf{P}[k]\right)^{-1}\mathbf{Q}[k]\mathbf{P}[k]\right)\\
    =&\left(\mathbf{I}-\mathbf{P}[k]\right)\left(\mathbf{I}-\mathbf{Q}[k]\mathbf{P}[k]\right)^{-1}.
\end{split}
\end{equation}
\end{subequations}

\begin{theorem}\label{theorem1}
 The traffic dynamics  \eqref{traffic dynamic state vector} is bounded-input bounded-output (BIBO) stable, when the following premises are satisfied:
\begin{enumerate}
    \item Vehicles can only enter the NOIR through an inlet road element and exit from the NOIR through an outlet road element.
    \item The out-neighbors of an inlet road element are all interior road elements, i.e. if $i\in \mathcal{V}_{in}$, then, $\mathcal{O}_i\subset \mathcal{V}_I$.
    \item Vehicles entering the NOIR through an inlet road element departs the NOIR within a finite time period.
    \item Every road element has at least one in-neighbor element or out-neighbor element, i.e., no road element is isolated in the NOIR $\left(\bigwedge_{i\in\mathcal{V}}\Box\left(\mathcal{I}_{i}\bigcup\mathcal{O}_{i}\neq\emptyset\right)\right)$.
% $\left(\mathcal{I}_{i}\bigcup\mathcal{O}_{i}\neq\emptyset$, if $i\in\mathcal{V} \right)$.
\end{enumerate}
\end{theorem}
{

}

\section{Traffic Congestion Control}\label{section 5}
% As it is defined in the section \label{section 2}, we achieve the objective of traffic management by controlling the boundary variables of the NOIR, i.e., the external traffic flow $\mathbf{s}[k]$. 
We use MPC to determine the boundary control $\mathbf{s}[k]$ at every discrete time $k$ by solving a quadratic programming problem with quadratic costs and linear constraints imposing the feasibility conditions into management of traffic coordination. To this end, according to Eq. \eqref{traffic dynamic state vector}, we predict the traffic dynamics within the next $N_{\tau}$ sampling times by obtaining the following finite-horizon predictive model: 
\begin{equation}\label{iterative model}
    \mathbf{X}[k]=\mathbf{G}[k]\mathbf{x}[k]+\mathbf{H}[k]\mathbf{U}[k]
\end{equation}
where
\begin{subequations}\label{state iternative model}
\begin{equation}\label{Vector X}
    \mathbf{X}[k]=\begin{bmatrix}
    \mathbf{x}[k+1]\\
    \vdots\\
    \mathbf{x}[k+N_\tau]
    \end{bmatrix}
    \in \mathbb{R}^{\left(N-N_{out}\right)N_{\tau}\times 1},
\end{equation}
\begin{equation}\label{MatrixG}
    \mathbf{G}[k]=\begin{bmatrix}
    \mathbf{A}[k]\\
    \vdots\\
    \mathbf{A}^{N_\tau}[k]
    \end{bmatrix}
    \in \mathbb{R}^{\left(N-N_{out}\right)N_{\tau}\times \left(N-N_{out}\right)},
\end{equation}
\begin{equation}
    \resizebox{0.99\hsize}{!}{%
    $
    \mathbf{H}[k]=\begin{bmatrix}
    \mathbf{B}[k] & 0 & 0 &\cdots & 0\\
    \mathbf{A}[k]\mathbf{B}[k] & \mathbf{B}[k] & 0 & \cdots & 0\\
    \mathbf{A}^2[k]\mathbf{B}[k] & \mathbf{A}[k]\mathbf{B}[k] & \mathbf{B}[k] & \cdots & 0\\
    \vdots & \vdots & \vdots & &\vdots \\
    \mathbf{A}^{N_\tau-1}[k]\mathbf{B}[k] & \mathbf{A}^{N_\tau-2}[k]\mathbf{B}[k] & \mathbf{A}^{N_\tau-3}[k]\mathbf{B}[k] & \cdots & \mathbf{B}[k]
    \end{bmatrix},
    % \in \mathbb{R}^{\left(N_\tau N\right)\times \left(N_\tau N_{out}\right)}
    $
    }
\end{equation}
\begin{equation}
    \mathbf{x}[k]=\begin{bmatrix}
    \rho_1[k]\\
    \vdots\\
    \rho_N[k]\\
    \end{bmatrix}
    \in \mathbb{R}^{\left(N-N_{out}\right)\times 1},
    \end{equation}
\begin{equation}
    \mathbf{U}[k]=\begin{bmatrix}
    \mathbf{s}[k]\\
    \vdots\\
    \mathbf{s}[k+N_\tau-1]\\
    \end{bmatrix}
    \in \mathbb{R}^{\left(N_{out}N_{\tau}\right)\times 1}.
\end{equation}
\end{subequations}

The cost function $\mathrm{C}$, previously defined in  \eqref{cost function}, can be rewritten as follows:
\begin{equation}\label{cost function state}
\begin{split}
\mathrm{C}\left(\mathbf{U}[k]\right)=&{1\over 2} \left(\mathbf{U}^\mathsf{T}[k]\mathbf{U}[k]+\beta\mathbf{X}^\mathsf{T}[k]\mathbf{X}[k]\right)\\
=&{1\over 2}\mathbf{U}^\mathsf{T}[k]\mathbf{W}_1[k]\mathbf{U}[k]+\mathbf{W}_2^\mathsf{T}[k]\mathbf{U}[k]+\mathbf{W}_3[k],
\end{split}
\end{equation}
where 
\begin{subequations}
    \begin{equation}
    \mathbf{W}_1[k]=\mathbf{I}_{N_{out}N_{\tau}}+\beta\mathbf{H}^\mathsf{T}[k]\mathbf{H}[k],
    \end{equation}
    \begin{equation}
    \mathbf{W}_2^\mathsf{T}[k]=\beta\mathbf{x}^\mathsf{T}[k]\mathbf{G}^\mathsf{T}[k]\mathbf{H}[k],
    \end{equation}
    \begin{equation}
    \mathbf{W}_3[k]={1\over 2}\beta\mathbf{x}^\mathsf{T}[k]\mathbf{G}^\mathsf{T}[k]\mathbf{G}[k]\mathbf{x}[k].
    \end{equation}
\end{subequations}

Note that $\mathbf{W}_3[k]$ can be removed from cost function  \eqref{cost function state} since $\mathbf{W}_3[k]$ depends on $\mathbf{x}[k]$ at every discrete time $k$, but it is independent of $\mathbf{U}[k]$. Therefore, 
\begin{equation}\label{cost function final}
    \mathrm{C}'={1\over 2}\mathbf{U}^\mathsf{T}[k]\mathbf{W}_1[k]\mathbf{U}[k]+\mathbf{W}_2[k]^\mathsf{T}\mathbf{U}[k]
\end{equation}
can be defined as the cost function of traffic coordination, and the optimal control variable
\begin{equation}
    \mathbf{s}^*[k]=\begin{bmatrix}
    \mathbf{I}_{N_{out}}&\mathbf{0}_{N_{out}\times N_{out}\left(N_\tau-1\right)}
    \end{bmatrix}
\end{equation}
is assigned by determining $\mathbf{U}^*[k]$ as the solution of the following optimization problem:
% To solve the optimal solution of this traffic dynamics equation at discrete sampling time $k$ within the next $N_{\tau}$ steps, 
% Furthermore, we can rewrite the traffic cost function  \eqref{cost function} and the traffic feasibility conditions  \eqref{State Feasibility Condition},  \eqref{Input Feasibility Condition}, and  \eqref{Input Optional Condition} as follows:
\[
\min \mathrm{C}'=\min~\left({1\over 2}\mathbf{U}^\mathsf{T}[k]\mathbf{W}_1[k]\mathbf{U}[k]+\mathbf{W}_2^\mathsf{T}[k]\mathbf{U}[k]\right)
 \]
subject to
\begin{subequations}\label{feasibility constrains}
\begin{equation}\label{inequality condition}
    \mathbf{W}_4[k]\mathbf{U}[k]+\mathbf{W}_5[k]\leq \mathbf{0},
\end{equation}
\begin{equation}\label{equality condition equation}
    \mathbf{W}_6[k]\mathbf{U}[k]+\mathbf{W}_7[k]= \mathbf{0},
\end{equation}
\end{subequations}
where
\begin{subequations}
\begin{equation}
\mathbf{W}_4[k]=
    \begin{bmatrix}
    -\mathbf{I}_{N_{out}N_\tau }\\
    % \mathbf{H}[k]\\
    \mathbf{H}[k]\\
    -\mathbf{H}[k]
    \end{bmatrix},
\end{equation}
\begin{equation}
\mathbf{W}_5[k]=
    \begin{bmatrix}
    \mathbf{0}_{N_{out}N_\tau \times 1}\\
    -\mathbf{1}_{\left(N-N_{out}\right) N_\tau\times 1}\otimes\mathbf{x}_{\mathrm{max}}+
    \mathbf{G}[k]\mathbf{x}[k]\\
    -\mathbf{G}[k]\mathbf{x}[k]
    % {\color{black}\mathbf{G}[k]\mathbf{x}[k]}
    \end{bmatrix},
\end{equation}
\begin{equation}
\mathbf{W}_6[k]=
    % \begin{bmatrix}
    \mathbf{I}_{N_\tau}\otimes\mathbf{1}_{1\times N_\mathrm{out}}
    % {\color{black}\mathbf{G}[k]\mathbf{x}[k]}
    % \end{bmatrix}
    ,
\end{equation}
\begin{equation}\label{equality condition}
\mathbf{W}_7[k]=
    % \begin{bmatrix}
    -d_0\mathbf{1}_{N_\tau\times 1}.
    % \end{bmatrix}.
\end{equation}
\end{subequations}

\begin{theorem}\label{TheoremNon-negative}
The traffic density at every interior road element and discrete time $k$ satisfies the inequality equation 
\begin{equation}
    \rho_{i}[k]\geq0,\qquad  \forall i\in\mathcal{V}_{I}\cup k\geq1,
\end{equation}
if conditions
\begin{subequations}\label{Theorem1conditions}
\begin{equation}
    \rho_{i}[1]\geq 0,\qquad \forall i\in\mathcal{V}_{I},
\end{equation}
\begin{equation}
    u_j[k]\geq 0,\qquad \forall j\in\mathcal{V}_{in},
\end{equation}
\begin{equation}
   v_n[k]\geq 0,\qquad \forall n\in\mathcal{V}_{out},
\end{equation}
\end{subequations}
% \begin{equation}
%     \begin{split}
%     &\rho_{i}[0]\geq 0,\qquad \forall i\in\mathcal{V}_{I}\\
%     &u_j[k]\geq 0,\qquad \forall j\in\mathcal{V}_{in}\\
%     &v_n[k]\geq 0,\qquad \forall n\in\mathcal{V}_{out}\\
%     \end{split}
% \end{equation}
hold.
\end{theorem}
\textbf{Proof:} See the proof in  \cite{liu2021conservation}.

% \textbf{Proof:} The Eq. \eqref{traffic dynamics interior node} describes the traffic dynamics of every interior road element $i\in\mathcal{V}_{I}$ in the NOIR. Per Assumption   \ref{assumption 1}, the flow probability $p_{i}[k]$ is defined in the interval $[0,1)$ for every interior road element and sampling time $k$ which implies that the term $1-p_{i}[k]$ in Eq. \eqref{traffic dynamics interior node} is always greater than 0. Moreover, the fraction probability $q_{i,j}[k]$ is also defined as a non-negative value in the interval $[0,1]$ at every sampling time $k$. Therefore, if conditions that the external traffic inflow $u_{j}[k]\geq0$ for all inlet road elements $j\in\mathcal{V}_{in}$, the external traffic inflow $v_{n}[k]\geq0$ for all outlet road elements $n\in\mathcal{V}_{out}$ and the initial traffic density $\rho_{i}[0]\geq 0$ at every road element $i\in\mathcal{V}_{I}$ are also satisfied, then the right hand side of Eq. \eqref{traffic dynamics interior node} must greater than 0 for every interior road element, i.e., $\rho_{i}[k]\geq0, \forall i\in\mathcal{V}_{I}\cup k\geq1$.
\begin{remark}
Per Theorem \ref{TheoremNon-negative}, it is ensured that the traffic density is non-negative at every discrete time $k$,  if conditions \eqref{Theorem1conditions} hold. Therefore, constraint Eq.  \eqref{inequality condition} simplifies to
\begin{equation}\label{inequality constrain sum}
    \resizebox{0.99\hsize}{!}{%
    $
    \begin{bmatrix}
    -\mathbf{I}_{N_\tau N_{out}}\\
    \mathbf{H}[k]
    \end{bmatrix}\mathbf{U}[k]+
    \begin{bmatrix}
    \mathbf{0}_{N_\tau N_{out}\times 1}\\
    -\mathbf{1}_{\left(N-N_{out}\right) N_\tau\times 1}\otimes\mathbf{x}_{\mathrm{max}}+
    \mathbf{G}[k]\mathbf{x}[k]
    % {\color{black}\mathbf{G}[k]\mathbf{x}[k]}
    \end{bmatrix}
    \leq0.
    $}
\end{equation}
\end{remark}

\section{Simulation Results}\label{section 6}
In this section, we illustrate the simulation results of the traffic control based on a part of real street map of the Philadelphia Center City which is composed of 259 road elements (See Fig.  \ref{fig:Philadelphia Center}). In order to denote and manage the road elements in the map conveniently, we assign specific indices for every road element in the NOIR. According to the road types, the road elements set $\mathcal{V}=\{1,\cdots,259\}$ is partitioned to $\mathcal{V}_{in}=\{1,\cdots,20\}$, $\mathcal{V}_{out}=\{21,\cdots,42\}$, and $\mathcal{V}_{I}=\{43,\cdots,259\}$.
% This section presents the simulation results of traffic coordination modeling and traffic congestion congestion control in an NOIR representing  Center City, Philadelphia with the map shown in
% Fig.  \ref{fig:Philadelphia Center}.  
% \begin{figure}
%     \centering
%     \includegraphics[width=0.47\textwidth]{paper pics/PhilaMap_with_idx.png}
%     \caption{Caption}
%     \label{fig:my_label}
% \end{figure}
% The selected area consists of $259$ road elements where the road index numbers are shown in Fig.   \ref{fig:Philadelphia Center}. {\color{black}We process the} map data {\color{black}generated by}  SUMO {\color{black}and obtain} graph $\mathcal{G}\left(\mathcal{V},\mathcal{E}\right)$. Set $\mathcal{V}=\{1,\cdots,259\}$ can  $\mathcal{V}=\mathcal{V}_{in}\bigcup \mathcal{V}_{out}\bigcup \mathcal{V}_{I}$ and $\mathcal{V}_{in}=\{1,\cdots,20\}, \mathcal{V}_{out}=\{21,\cdots,42\}, \mathcal{V}_{I}=\{43,\cdots,259\}$.

We randomly assign the initial traffic density $\rho_{i}[0]$ at every road element $i\in\mathcal{V}$ in the NOIR and run the simulation for $300$ time steps, where sampling time $k$ represents the continuous time interval $\left[t_k,t_{k+1}\right)$ and $\Delta t=t_{k+1}-t_k$ is constant at every discrete time $k$.
% , to observe the variation of some traffic characteristic parameters, such as traffic flow and traffic density. And 
At each sampling time $k$, 
the flow probability  matrix $\mathbf{P}[k]$ and fraction probability matrix $\mathbf{Q}[k]$ are randomly generated.

For simulation, we ignore the inter-vehicle distance and assume that the length of vehicle $l_{veh}$ is equal to $4.5m$ and obtain the maximum traffic density $\rho_{i,max}$ for every interior road element $i\in\mathcal{V}_{I}$ by
\begin{equation}
    \rho_{i,max}=\frac{n_{i,lane}\times l_{i}}{l_{veh}},
\end{equation} 
% Moreover, in order to constrain the traffic density at every interior road element such that it less than the maximum capacity of the road, we assume that for all interior road elements, the length of vehicle $l_{veh}$ is equal to 4.5m. Then, the maximum traffic density $\rho_{i,max}$ for every interior road element $i\in\mathcal{V}_{I}$ can be determined by:
% \begin{equation}
%     \rho_{i,max}=\frac{n_{i,lane}*l_{i}}{l_{veh}}
% \end{equation}
where $l_{i}$ is the length of the road element $i$ obtained from the street map data and $n_{i,lane}$ is the number of lanes in road element $i\in \mathcal{V}_I$. Furthermore, we assume that $400$ cars can cross the boundary of the NOIR during the interval $[t_{k},t_{k+1})$ at every sampling time $k$. Therefore, $d_{0}=400$ is used in equality constraint \eqref{equality condition}.

% Moreover, to present the effect from the weigh parameter $\beta$ on the optimal solution, 
The simulation is run under different scaling parameters $\beta=0$, $\beta=0.5$ and  $\beta=1$ that are  used to weigh the cost function  \eqref{cost function state}. The results are presented for two boundary inlet, boundary outlet, and interior road elements with index numbers and locations presented in Table   \ref{exapmle chosen roads}. 
\begin{table}[h]
    \centering
    \begin{tabular}{|c|c|m{3.2cm}|}
    \hline
    \textbf{Road Type} & \textbf{Road Element Index} & \textbf{Name and Location}\\
    \hline
    Inlet & 9 & Chestnut St. between S19th St. and S20th St.\\
    \hline
    Inlet & 12 & Sansom St. between S9th St. and S10th St.\\
    \hline
    Outlet & 28 & Filbert St. between S9th St. and S10th St.\\
    \hline
    Outlet & 33 & Arch St. between N19th St. and N20th St.\\
    \hline
    Interior & 150 & Market St. between N16th St. and N17th St.\\
    \hline
    Interior & 239 & Walnut St. between S15th St. and S16th St.\\
    \hline
    \end{tabular}
    \caption{Example road elements in NOIR}
    \label{exapmle chosen roads}
\end{table}

\begin{figure}
    \centering
    \includegraphics[width=0.48\textwidth]{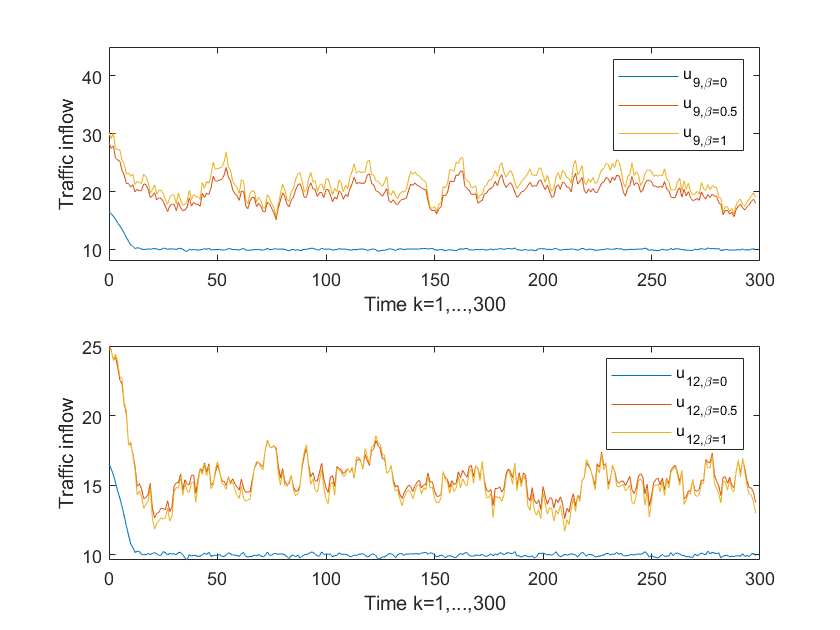}
    \caption{External traffic inflows of inlet road elements under different $\beta$ values}
    \label{fig:inlet node density and inflow}
\end{figure}

\begin{figure}
    \centering
    \includegraphics[width=0.48\textwidth]{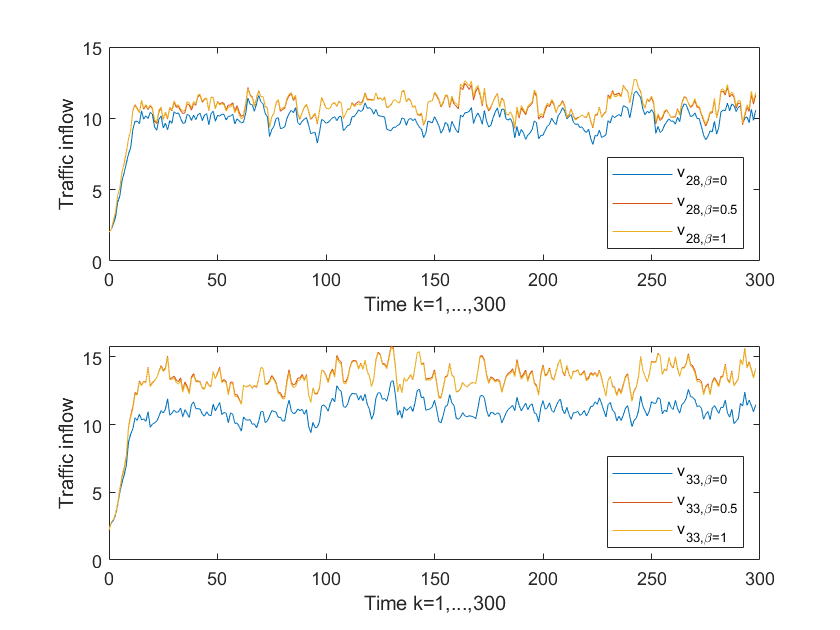}
    \vspace{-.3cm}
    \caption{External traffic outflows of outlet road elements under different $\beta$ values}
    \label{fig:outlet node density and inflow}
\end{figure}

We illustrate the variation of the optimal external traffic flow $s_{i}$ at the example road elements under different values of $\beta$ throughout the whole simulation time in Figs.   \ref{fig:inlet node density and inflow} and Fig.  \ref{fig:outlet node density and inflow}.   
% {\color{red}By incorporating the scaling parameter $\beta$ in the cost function, the optimal solution at every boundary element can be allocated under the consideration of the number of existing cars in the NOIR.}
It is observed that the external traffic flow of the inlet road elements $u_{i}$ and outlet road elements $v_{i}$ hold similar properties at different values of $\beta$, that is: the external traffic inflow and outflow reach the steady state condition after about $20$ sampling times. For $\beta=0$, the weight matrix $\mathbf{W}_1[k]$ is diagonal at every discrete time $k$, but $\mathbf{W}_1[k]$ is not diagonal and $\mathbf{W}_2^\mathsf{T}[k]$ also not equals to $\mathbf{0}$ when $\beta>0$ is selected. Therefore, we observe that the variations of external flow are different when $\beta=0$ is selected. In addition, it is observed that that the plots have large fluctuations when $\beta>0$  due to the variation of the number of existing cars in the interior road elements.
% The reason for this phenomenon is that, different values of $\beta$ influence the optimization model. Note that, the cost function   \ref{cost function final} when $\beta=1$ is a standard quadratic equation with respect to all control variables $s_{i}$, i.e.,:
% \begin{equation}
%     \mathrm{C}=a_1*s_1^{2}+a_2*s_2^{2}+\cdots+a_n*s_n^{2},
% \end{equation}
% where $a_1,\cdots,a_n$ are some constant values. However, when $\beta$ is equal to 0.5 or 1, the cost function will become the sum of a quadratic equation and a polynomial:
% \begin{equation}
%     \mathrm{C}=a_1*s_1^{2}+a_2*s_2^{2}+\cdots+a_n*s_n^{2}+b_1*s_1+b_2*s_2+b_n*s_n,
% \end{equation}
% where $b_1,\cdots,b_n$ are constant values of the polynomial. Therefore, optimization under same constrains with changed cost functions lead to this differences.   
% However, it can be observed that the existence of $\beta$ has a significant impact on the change trend of the traffic flow. 

\begin{figure}
    \centering
    \includegraphics[width=0.48\textwidth]{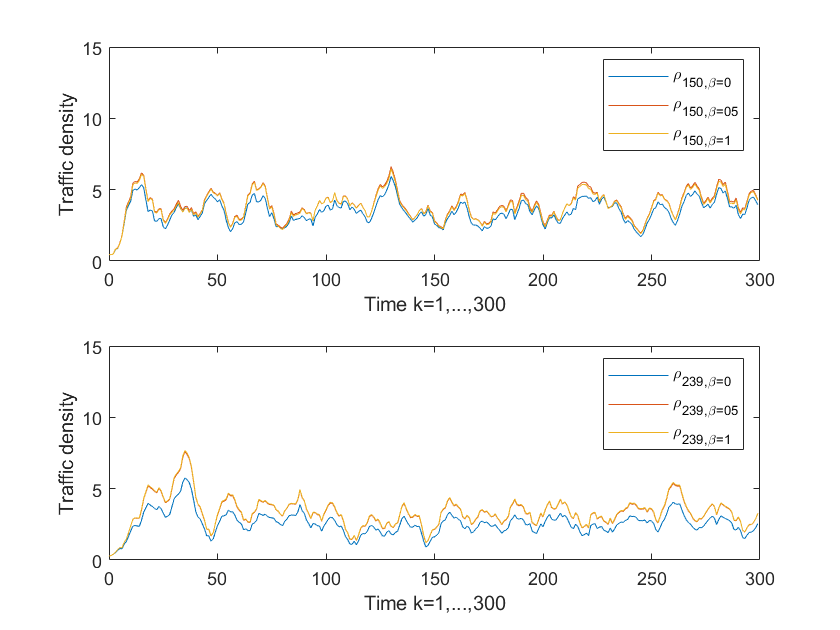}
    \vspace{-.3cm}
    \caption{Traffic densities of interior road elements under different $\beta$ values}
    \label{fig:interior traffic density}
\end{figure}

Fig.  \ref{fig:interior traffic density}, plots variations of $\rho_{150}[k]$ and $\rho_{239}[k]$ at different sampling times $k$ ($k\in \left\{1,\cdots,300\right\}$). Although, variation curves of traffic densities at interior road elements have a large fluctuation, we could still observe an stable tendency after a certain period of time.

\begin{figure}
    \centering
    \includegraphics[width=0.48\textwidth]{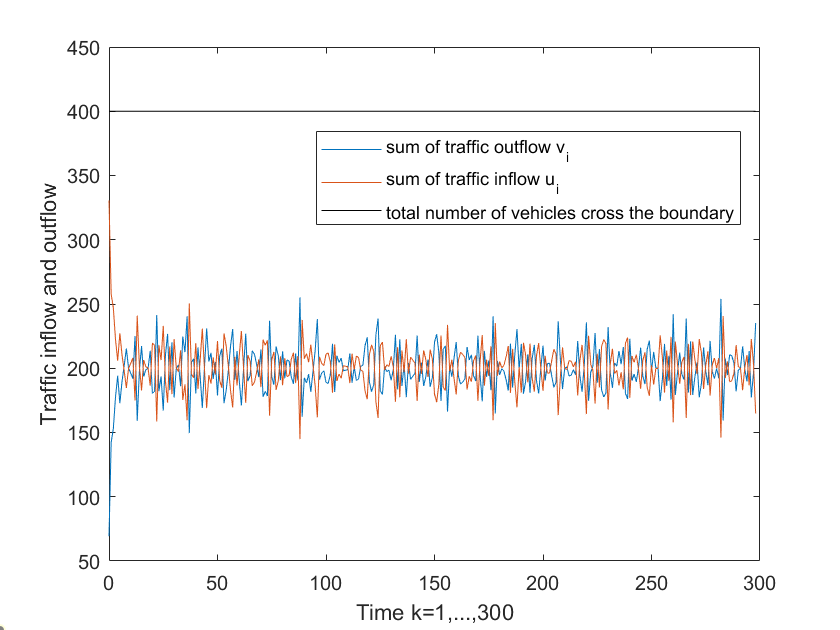}
    \vspace{-.3cm}
    \caption{External traffic inflow and outflow of the NOIR under $\beta=0$}
    \label{fig:total number of inflow and outflow beta0}
\end{figure}

\begin{figure}
    \centering
    \includegraphics[width=0.48\textwidth]{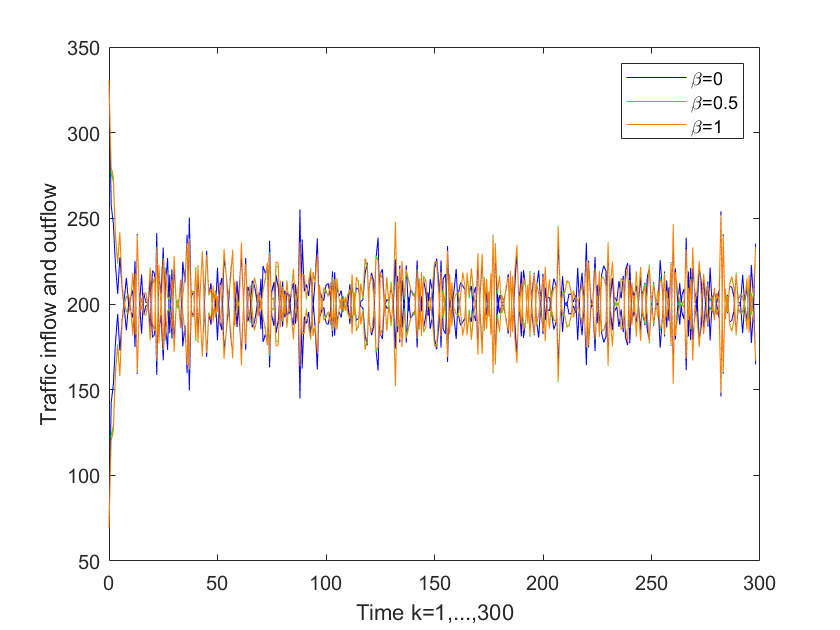}
    \vspace{-.3cm}
    \caption{External traffic inflow and outflow of the NOIR under different $\beta$ values}
    \label{fig:total number of inflow and outflow}
\end{figure}

The number of vehicles entering and existing the NOIR  during the whole simulation time are plotted in the Fig.  \ref{fig:total number of inflow and outflow beta0} for $\beta=0$. Consistent with the constraint Eq.  \eqref{equality condition equation}, the sum of vehicles crossing the border of the NOIR is equal to $d_0=400$ at every sampling time $k$. Starting from a larger value, the amount of external traffic inflow $u_{i}$ gradually decreases to a steady-state value at $200$. It is seen that the external traffic outflow $v_{i}$ symmetrically increases from a small value and reaches the steady state value at $200$. The simulation results of external traffic flows for $\beta=0.5$ and $\beta=1$ shown in Fig.  \ref{fig:total number of inflow and outflow} illustrate a similar trend as when $\beta=0$. Note the equilibrium (steady-state) condition, which implies that the traffic entering the NOIR is equal to the traffic existing from the NOIR, is first observed at $k\geq20$ where
\begin{equation}
    \forall k\geq20,\sum_{i\in\mathcal{V}_{in}}u_i[k]=\sum_{i\in\mathcal{V}_{out}}v_i[k]\cong200
\end{equation}

% The net inflow and outflow of the NOIR are plotted versus discrete time $k$ in Fig.   \ref{fig:total number of inflow and outflow}. The red plot illustrated the total number of cars entering the NOIR through all inlet nodes defined by set $\in\mathcal{V}_{in}$ and the blue plot shows the total number of cars leaving the NOIR though outlet nodes defined by set $\mathcal{V}_{out}$. The black line indicates the sum of traffic inflow and outflow at every discrete time $k$ which will maintain a constant value at $u_{0}=100$ during the whole simulation. It is observed that
% \begin{equation}
%     \forall k\geq30,\sum_{i\in\mathcal{V}_{in}}u_i[k]=\sum_{i\in\mathcal{V}_{out}}v_i[k]\cong50
% \end{equation}
% which implies that the total number of cars entering the NOIR is approximately equal to the total number of cars departing the NOIR for $k\geq30$.  
% \vspace{-.3cm}
\section{Conclusion}\label{section 7}
This paper intruduces a physics-inspired approach based on the mass conservation law to model the traffic dynamics and implements the MPC method to control the boundary traffic inflow and outflow, so that the traffic congestion can be alleviated. Comparing with our previous research, the traffic dynamics model in this paper is more realistic. Simulation applied in a area of Philadelphia Center City demonstrates that the proposed traffic model and control approach can achieve the objective of traffic congestion alleviation successfully through controlling the boundary traffic flow. Our further research will focus on the integration of the Makrov decision process (MDP) with the traffic dynamics model to control the traffic more efficiently and intelligently. 

% The authors gratefully thank Dr. Sergey Nersesov for the useful comments on this paper.

% This paper introduced a conservation-based approach to learn the traffic network dynamics and alleviate the traffic congestion. We applied the mass conservation law to model traffic coordination by a time-varying stochastic process where the  real map data was used to define the traffic network. We offered an MPC control to manage traffic congestion by controlling the inflow and outflow at the boundary of the NOIR. The simulation results show that our proposed approach can effectively control traffic congestion through optimizing the boundary inflow and outflow. 
% Applying both boundary inlet road elements and outlet road elements as the control variables, the research on the network dynamics can have more possibilities. Furthermore, the integration with the MPC approach makes consistent implementation of optimal boundary control variables into the network model possible. The reasonable results shown in the simulation section illustrate that this proposed approach demonstrates a convincing ability to effectively control the urban scale network. 
% In our future work, we plan to obtain the traffic dynamics based on real traffic data and control congestion through the boundary ramp meters and traffic signals, situated at road intersections.
% \vspace{-.3cm}
\section{Acknowledgement}
This work has been supported by the Department of Mechanical Engineering at Villanova University. The authors would like to gratefully acknowledge Dr. Sergey Nersesov for the useful comments on this paper, and the Mechanical Engineering PhD fellowship provided to Xun Liu which was made possible by a generous gift from Dr. Yongping Gu and Fei Gu.
\vspace{-.3cm}

\bibliographystyle{IEEEtran}
\bibliography{reference}

\begin{thebibliography}{10}
\providecommand{\url}[1]{#1}
\csname url@samestyle\endcsname
\providecommand{\newblock}{\relax}
\providecommand{\bibinfo}[2]{#2}
\providecommand{\BIBentrySTDinterwordspacing}{\spaceskip=0pt\relax}
\providecommand{\BIBentryALTinterwordstretchfactor}{4}
\providecommand{\BIBentryALTinterwordspacing}{\spaceskip=\fontdimen2\font plus
\BIBentryALTinterwordstretchfactor\fontdimen3\font minus
  \fontdimen4\font\relax}
\providecommand{\BIBforeignlanguage}[2]{{%
\expandafter\ifx\csname l@#1\endcsname\relax
\typeout{** WARNING: IEEEtran.bst: No hyphenation pattern has been}%
\typeout{** loaded for the language `#1'. Using the pattern for}%
\typeout{** the default language instead.}%
\else
\language=\csname l@#1\endcsname
\fi
#2}}
\providecommand{\BIBdecl}{\relax}
\BIBdecl

\bibitem{chin2011impact}
H.~Chin and M.~Rahman, ``An impact evaluation of traffic congestion on
  ecology,'' \emph{Planning Studies and Practice}, vol.~3, pp. 32--44, 2011.

\bibitem{liang2013road}
L.~Ye, Y.~Hui, and D.~Yang, ``Road traffic congestion measurement considering
  impacts on travelers,'' \emph{Journal of Modern Transportation}, vol.~21, pp.
  28--39, 2013.

\bibitem{annan2015traffic}
J.~Annan, J.~Mensah, and N.~Boso, ``Traffic congestion impact on energy
  consumption and workforce productivity:empirical evidence from a developing
  country,'' \emph{Archives of Business Research}, vol.~3, pp. 40--54, 2015.

\bibitem{muneera2018economic}
C.~P. Muneera and K.~Karuppanagounder, ``Economic impact of traffic congestion-
  estimation and challenges,'' \emph{European Transport / Trasporti Europei},
  vol.~68, 2018.

\bibitem{margaret2004impact}
M.~O’Mahony and H.~Finlay, ``Impact of traffic congestion on trade and
  strategies for mitigation,'' \emph{Transportation Research Board}, vol. 1873,
  pp. 25--34, 2004.

\bibitem{geroliminis2008existence}
\BIBentryALTinterwordspacing
N.~Geroliminis and C.~F. Daganzo, ``Existence of urban-scale macroscopic
  fundamental diagrams: Some experimental findings,'' \emph{Transportation
  Research Part B: Methodological}, vol.~42, no.~9, pp. 759 -- 770, 2008.
  [Online]. Available:
  \url{http://www.sciencedirect.com/science/article/pii/S0191261508000180}
\BIBentrySTDinterwordspacing

\bibitem{xu2013traffic}
F.~Xu, Z.~He, Z.~Sha, W.~Sun, and L.~Zhuang, ``Traffic state evaluation based
  on macroscopic fundamental diagram of urban road network,'' \emph{Procedia
  Social and Behavioral Sciences}, vol.~96, pp. 480--489, 2013.

\bibitem{sirmatel2017integration}
I.~I. Sirmatel and N.~Geroliminis, ``Integration of perimeter control and route
  guidance in large-scale urban networks via model predictive control,'' in
  \emph{Transportation Research Board 96th Annual Meeting}.\hskip 1em plus
  0.5em minus 0.4em\relax Transportation Research Board, 2017, p. 13p.

\bibitem{shao2018distributed}
P.~Shao, L.~Wang, W.~Qian, Q.-G. Wang, and X.-H. Yang, ``A distributed traffic
  control strategy based on cell-transmission model,'' \emph{IEEE Access},
  vol.~6, pp. 10\,771--10\,778, 2018.

\bibitem{munoz2003traffic}
L.~Munoz, X.~Sun, R.~Horowitz, and L.~Alvarez-Icaza, ``Traffic density
  estimation with the cell transmission model,'' vol.~5, 07 2003, pp. 3750 --
  3755.

\bibitem{yin2017offblock}
S.~Yin, L.~Yang, and K.~Han, ``Off-block flow optimisation based on cell
  transmission model.''\hskip 1em plus 0.5em minus 0.4em\relax DASC, 04 2017.

\bibitem{feldman2002cell}
O.~Feldman and M.~Maher, ``A cell transmission model applied to the
  optimisation of traffic signals,'' 01 2002.

\bibitem{yang2017fundamental}
L.~Yang, S.~Yin, K.~Han, J.~Haddadc, and M.~Hu, ``Fundamental diagrams of
  airport surface traffic: Models and applications,'' \emph{Transportation
  Research Part B: Methodological}, vol. 106, pp. 29--51, 2017.

\bibitem{yang2019network}
L.~Yang, S.~Yin, and M.~Hu, ``Network flow dynamics modeling and analysis of
  arrival traffic in terminal airspace,'' \emph{IEEE Access}, vol.~7, pp.
  73\,993--74\,016, 06 2019.

\bibitem{liu2021conservation}
X.~Liu and H.~Rastgoftar, ``Conservation-based modeling and boundary control of
  congestion with an application to traffic management in center city
  philadelphia,'' \emph{arXiv preprint arXiv:2102.00552}, 2021.

\bibitem{rastgoftar2019integrative}
H.~Rastgoftar and E.~Atkins, ``An integrative data-driven physics-inspired
  approach to traffic congestion control,'' 2019.

\bibitem{zhang2021optimization}
Y.~Zhang and R.~Su, ``An optimization model and traffic light control scheme
  for heterogeneous traffic systems,'' \emph{Transportation Research Part C:
  Emerging Technologies}, vol. 124, p. 102911, 03 2021.

\bibitem{ba2016ondistributed}
Q.~Ba and K.~Savla, ``On distributed computation of optimal control of traffic
  flow over networks,'' in \emph{54th Annual Allerton Conference on
  Communication, Control, and Computing}.\hskip 1em plus 0.5em minus
  0.4em\relax IEEE, 2016, pp. 1102--1109.

\bibitem{rastgoftar2021physicsbased}
H.~Rastgoftar and J.-B. Jeannin, ``A physics-based finite-state abstraction for
  traffic congestion control,'' 2021.

\bibitem{rastgoftar2020resilient}
H.~Rastgoftar and A.~Girard, ``Resilient physics-based traffic congestion
  control,'' 07 2020, pp. 4120--4125.

\bibitem{lin2011fast}
S.~Lin, B.~De~Schutter, Y.~Xi, and H.~Hellendoorn, ``Fast model predictive
  control for urban road networks via milp,'' \emph{Intelligent Transportation
  Systems, IEEE Transactions on}, vol.~12, pp. 846 -- 856, 10 2011.

\bibitem{li2019model-free}
Z.~Li, S.~Jin, C.~Xu, and J.~Li, ``Model-free adaptive predictive control for
  an urban road traffic network via perimeter control,'' \emph{IEEE Access},
  vol.~7, pp. 172\,489--172\,495, 11 2019.

\bibitem{abdulhai2003reinforcementlearning}
B.~Abdulhai, R.~Pringle, and G.~Karakoulas, ``Reinforcement learning for true
  adaptive traffic signal control,'' \emph{Journal of Transportation
  Engineering}, vol. 129, 05 2003.

\bibitem{prashanth2011reinforcementlearning}
P.~L.A. and S.~Bhatnagar, ``Reinforcement learning with function approximation
  for traffic signal control,'' \emph{Intelligent Transportation Systems, IEEE
  Transactions on}, vol.~12, pp. 412 -- 421, 07 2011.

\bibitem{lin2018efficient}
Y.~Lin, X.~Dai, L.~Li, and F.-Y. Wang, ``An efficient deep reinforcement
  learning model for urban traffic control,'' 2018.

\bibitem{greguric2020application}
M.~Gregurić, M.~Vujić, C.~Alexopoulos, and M.~Miletić, ``Application of deep
  reinforcement learning in traffic signal control: An overview and impact of
  open traffic data,'' \emph{Applied Sciences}, vol.~10, p. 4011, 06 2020.

\bibitem{oluyemi2019development}
O.~Adetoyi, ``Development of sugeno fuzzy controlled traffic system for y-road
  intersection – university of ibadan case study,'' 11 2019.

\bibitem{bhatia2021smart}
M.~Bhatia, A.~Aggarwal, and N.~Kumar, ``Smart traffic light system to control
  traffic congestion pjaee, 17 (9) (2020) smart traffic light system to control
  traffic congestion,'' \emph{PalArch's Journal of Archaeology of Egypt/
  Egyptology}, vol.~17, pp. 7093--7109, 01 2021.

\bibitem{zohar1991temporal}
A.~P. Zohar~Manna, \emph{The Temporal Logic of Reactive and Concurrent
  Systems}.\hskip 1em plus 0.5em minus 0.4em\relax Springer-Verlag New York,
  1992.

\bibitem{wongpiromsarn2009receding}
T.~Wongpiromsarn, U.~Topcu, and R.~Murray, ``Receding horizon temporal logic
  planning for dynamical systems,'' 12 2009, pp. 5997--6004.

\bibitem{koymans1990specifying}
R.~Koymans, ``Specifying real-time properties with metric temporal logic,''
  \emph{Real-Time Systems}, vol.~2, p. 255–299, 1990.

\bibitem{qu2009cooperative}
Z.~Qu, \emph{Cooperative Control of Dynamical Systems}.\hskip 1em plus 0.5em
  minus 0.4em\relax Springer-Verlag London, 2009.

\end{thebibliography}

\appendix
\textbf{Proof of Theorem \ref{theorem1}:}
Given the above definitions, matrix $\mathbf{Q}[k]\mathbf{P}[k]$ holds the following properties:
\begin{enumerate}
    \item{All entries of matrix $\mathbf{Q}[k]\mathbf{P}[k]$ are non-negative.}
    \item{Since no isolated interior road element exists in the NOIR, entries of column $i\in \{1,\cdots,N-N_{out}\}$ of  $\mathbf{Q}[k]\mathbf{P}[k]$ sum up to a positive value in the interval $(0,1)$.}
\end{enumerate}

Given the above characteristics, the spectral radius of matrix $\mathbf{Q}[k]\mathbf{P}[k]$ is less than $1$ at every sampling time $k$ \cite{qu2009cooperative}. Then, according to the Neumann series theorem, we can rewrite the matrix $\left(\mathbf{I}-\mathbf{Q}[k]\mathbf{P}[k]\right)^{-1}$ as follows:
\begin{equation}\label{neumann}
    \left(\mathbf{I}-\mathbf{Q}[k]\mathbf{P}[k]\right)^{-1}= \sum_{h=0}^\infty\left(\mathbf{Q}[k]\mathbf{P}[k]\right)^{h}.
\end{equation}
  Pre-multiplying both sides of  \eqref{neumann} by $\left(\mathbf{I}-\mathbf{Q}[k]\mathbf{P}[k]\right)$, we obtain
\begin{equation}\label{neumann_exp}
\begin{split}
    &\left(\mathbf{I}-\mathbf{Q}[k]\mathbf{P}[k]\right)\left(\mathbf{I}-\mathbf{Q}[k]\mathbf{P}[k]\right)^{-1}\\
    &=\sum_{h=0}^\infty\left(\mathbf{Q}[k]\mathbf{P}[k]\right)^{h}-\sum_{h=0}^\infty\left(\mathbf{Q}[k]\mathbf{P}[k]\right)^{h+1}\\
    % &=\mathbf{I}-\mathbf{Q}[k]\mathbf{P}[k
    % ]^{\infty+1}\\
    &=\lim_{h \to\infty}\left(\mathbf{I}-\left(\mathbf{Q}[k]\mathbf{P}[k]\right)^{h+1}\right)=\mathbf{I}.
\end{split}
\end{equation}
Since every entry of matrix $\mathbf{Q}[k]$ is non-negative and sum of the elements of $N_{out}-N_{in}$ columns of matrix $\mathbf{Q}[k]$ are less than $1$, the spectral radius of matrix $\mathbf{Q}[k]$ is less than $1$\cite{qu2009cooperative}. Then, the spectral radius of $\mathbf{Q}[k]\mathbf{P}[k]$ is less than the spectral radius of  matrix $\mathbf{P}[k]$. Therefore, the spectral radius of matrix $\mathbf{I}-\mathbf{P}[k]$ is less than the spectral radius of matrix $\mathbf{I}-\mathbf{Q}[k]\mathbf{P}[k]$ which in turn implies that the spectral radius of matrix $\mathbf{A}[k]=\left(\mathbf{I}-\mathbf{P}[k]\right)\left(\mathbf{I}-\mathbf{Q}[k]\mathbf{P}[k]\right)^{-1}$ is less than $1$ at every discrete time $k$. 

Now, we can rewrite the traffic dynamics  \eqref{traffic dynamic state vector} as:
\begin{equation}
\label{MAINNNNTheorem4}
\begin{split}
 \mathbf{x}[k+1]=&
 \mathbf{\Theta}_k
 \begin{bmatrix}
    \mathbf{x}[1]\\
    \mathbf{B}[1]\mathbf{s}[1]\\
    \vdots\\
    \mathbf{B}[k]\mathbf{s}[k]\\
    \end{bmatrix}
   \end{split}
\end{equation}
where 
\begin{subequations}
\begin{equation}
\mathbf{\Theta}_k=
    \begin{bmatrix}
 \mathbf{\Gamma}_{k}&\cdots&\mathbf{\Gamma}_1&\mathbf{\Gamma}_0
 \end{bmatrix}
 \end{equation}
 \begin{equation}\label{Gamma_h}
\mathbf{\Gamma}_{h}=\prod_{j=k-h+1}^k \mathbf{A}[j]
 \end{equation}
\end{subequations}
for $h=1,\cdots,k$, and $\mathbf{\Gamma}_0=\mathbf{I}_{N-N_{out}}\in \mathbb{R}^{\left(N-N_{out}\right)\times N-N_{out}}$ is an identity matrix. Because the initial traffic density vector $\mathbf{x}[1]$ and and control input vector $\mathbf{s}[k]$ are bounded, we can say:
\begin{subequations}
\begin{equation}
    \mathbf{x}[1]\leq z_\mathrm{max}\mathbf{1}_{N-N_{out}\times1}
\end{equation}
\begin{equation}
    \mathbf{B}\mathbf{s}[k]\leq z_\mathrm{max}\mathbf{1}_{N-N_{out}\times1},
\end{equation}
\end{subequations}
where $z_{\mathrm{max}}$ is a sufficiently large value. Moreover, as shown in Eq. \eqref{Gamma_h}, $\mathbf{\Gamma}_{h}$ is the product of matrices $\mathbf{A}[j]$. Therefore, we can draw the conclusion that the spectral radius of matrix $\mathbf{\Gamma}_h$ must also be less than $1$ at every discrete time $k$, since the spectral radius of matrix $\mathbf{A}[k]$ is less than upper bound $r_a<1$ at every sampling time $k$. Now, we calculate the norm of $||\mathbf{x}[k+1]||^2$ as follow:
\begin{equation}\label{boundedproof_spe}
\resizebox{0.99\hsize}{!}{%
    $
\begin{split}
    \mathbf{x}^\mathsf{T}\left[k+1\right]\mathbf{x}\left[k+1\right]&\leq z_{max}\mathbf{1}_{N-N_{out}\times1}^\mathsf{T}\left(\sum_{l=0}^k\sum_{h=0}^k\mathbf{\Gamma}_l^\mathsf{T}\mathbf{\Gamma}_h\right)z_{max}\mathbf{1}_{N-N_{out}\times1}\\
    &\leq z_{max}^2\left(N-N_{out}\right)\left(\sum_{l=0}^\infty r_{a}^l\right)\leq\dfrac{z_{max}^2\left(N-N_{out}\right)}{\left(1-r_{a}\right)}.
\end{split}
$
}
\end{equation}
which implies that $||\mathbf{x}[k+1]||^2$ is bounded at every discrete time $k$. Thus the BIBO stability of traffic dynamics  \eqref{traffic dynamic state vector} is proven.

\end{document}